# Hidden valley dynamics behind vanishing circular polarization in moiré excitons

*Yuto Urano[1,2], Lata Chouhan[1], Nurul Fariha Ahmad[1], Kenji Watanabe[3], Takashi Taniguchi[1], Daichi Kozawa[1,4], and Ryo Kitaura[1,2,*]*

[1] *Research Center for Materials Nanoarchitectonics, National Institute for Materials Science, 1-1 Namiki, Tsukuba, Ibaraki 305-0044, Japan*

[2] *Graduate School of Chemical Sciences and Engineering, Hokkaido University, 5, Kita 8 Nishi, Kita-ku, Sapporo-city, Hokkaido 060-8628, Japan*

[3] *Research Center for Electronic and Optical Materials, National Institute for Materials Science, 1-1 Namiki, Tsukuba, Ibaraki 305-0044, Japan*

[4] *Department of Materials Science, Institute of Pure and Applied Sciences, University of Tsukuba, 1-1-1 Tennodai, Tsukuba, Ibaraki 305-8573, Japan*

* Contact author: KITAURA.Ryo@nims.go.jp

**Abstract**

Optically addressable valley degrees of freedom in transition-metal dichalcogenide heterostructures provide a powerful platform for valleytronic and quantum-optical functionalities. In moiré superlattices, interlayer excitons inherit valley-contrasting optical selection rules while acquiring long lifetimes, electric dipoles, and site-dependent optical responses. However, because conventional measurements typically probe time-integrated valley polarization, the dynamical origin of vanishing polarization has remained elusive. Here, we show that a nearly zero steady-state valley polarization in electrically tunable moiré excitons does not necessarily indicate fast valley relaxation. Helicity-resolved time-resolved photoluminescence reveals a temporal crossing between $\sigma^+$ and $\sigma^-$ emission, indicating that helicity-opposite dynamical components coexist and compensate after time integration. A minimal two-channel model, representing A-like and B-like moiré emission channels with opposite optical selection rules and distinct effective decay/depolarization rates, reproduces the observed helicity crossing without invoking a single rapid valley relaxation process. Furthermore, the crossing time evolves systematically with electrostatic tuning, demonstrating that the hidden valley dynamics are electrically controllable. These results show that time-integrated circular polarization can give a false-negative indication of valley polarization in multichannel valley emitters.

## I. INTRODUCTION

Valley degrees of freedom in transition-metal dichalcogenide semiconductors can be initialized and read out optically through valley-contrasting circular selection rules, providing a direct route to valleytronic and quantum-optical functionalities [1–4]. In van der Waals heterobilayers, the formation of interlayer excitons further enriches this physics: electrons and holes reside in different layers, giving rise to long optical lifetimes, permanent out-of-plane electric dipoles, and strong electrostatic tunability [5–8]. When a moiré superlattice is formed, the excitonic landscape becomes spatially modulated, and different local atomic registries can host exciton states with distinct optical selection rules [9–15]. As quantum-dot-like states carrying a valley degree of freedom, moiré excitons offer an attractive platform for both strongly correlated moiré physics and optically addressable valley control [16–19].

The readout of the valley degree of freedom can be performed through measuring the degree of circular polarization (DCP) in steady-state photoluminescence (PL). A large DCP is commonly used as an experimental indicator of robust valley polarization or suppressed intervalley depolarization and vice versa[2,3,6,7,20,21]. A vanishing time-integrated DCP is, however, not, by itself, evidence for rapid valley depolarization. Because steady-state PL integrates over the full emission dynamics, helicity-opposite components with different effective decay or depolarization dynamics can compensate each other and yield an apparently unpolarized signal. This possibility is particularly relevant for moiré excitons, where A-like and B-like moiré sites can possess opposite optical selection rules and distinct local excitonic environments [10,15].

Here, we use helicity- and time-resolved PL to reveal hidden valley dynamics in electrically tunable $WSe_2/WS_2$ moiré excitons [13,15,19–21]. In regimes where the time-integrated circular polarization nearly vanishes, the $\sigma^+$ and $\sigma^-$ emission components exhibit a clear temporal crossing, showing that the steady-state cancellation does not necessarily correspond to a featureless depolarized state. This behavior is captured by a minimal two-channel model in which different moiré emission channels have opposite optical selection rules and distinct effective relaxation rates [10,15]. The crossing time evolves systematically with gate voltage, demonstrating electrostatic control of the competing helicity dynamics [8,20]. Our results show that steady-state valley polarization alone is insufficient to determine valley relaxation dynamics in moiré excitons and establish time-resolved helicity measurements as a sensitive probe of competing moiré emission channels.

## II. RESULTS AND DISCUSSION

We first characterize the $WSe_2/WS_2$ heterostructure device used for helicity-resolved optical measurements. Figure 1(a) shows an optical micrograph of the dual-gated device, and Fig. 1(b) illustrates the device structure. The $WSe_2/WS_2$ heterobilayer is encapsulated by 10-nm-thick top hexagonal boron nitride (hBN) and 15-nm-thick bottom hBN layers and is electrostatically controlled by top and bottom graphite gates with voltages $V_t$ and $V_b$, respectively. Second harmonic generation (SHG) measurements were used to identify the crystallographic orientations of $WSe_2$ and $WS_2$ (Fig. 1(c)). The extracted armchair orientations are −0.43 ± 0.17° for $WS_2$ and −0.57 ± 0.10° for $WSe_2$, giving a relative twist angle $\theta$ = 0.14 ± 0.20°. Because armchair orientations alone do not distinguish between near-0° and near-60° stacking configurations, we also measured the SHG response in the heterobilayer region. The corresponding SHG data are shown in Fig. S1 of the Supplemental Material. The heterobilayer signal is larger than that of the individual monolayers, indicating constructive interference of the SHG fields; we therefore assign the alignment to the near-0° (R-type) configuration rather than the near-60° (H-type) configuration, consistent with the moiré-exciton picture discussed below [13,15]. The low-temperature photoluminescence (PL) measured at 4 K with $V_t = V_b = 0$ shows a dominant interlayer-exciton emission band and no detectable intralayer-exciton emission (Fig. 1(d)). The approximately 100 meV width describes the envelope of several overlapping transitions rather than the linewidth of a single transition. The absence of detectable intralayer emission is consistent with ultrafast interlayer charge transfer enabled by strong interlayer coupling [22]. Consequently, the low-temperature PL is dominated almost exclusively by interlayer excitons and serves as the basis for the helicity-resolved measurements detailed below [13,15].

We then examine the steady-state circular polarization of the moiré emission. Figures 2(a)–2(c) show representative helicity-resolved PL spectra measured at three gate-field conditions. Depending on the electrostatic tuning, the time-integrated circular polarization changes sign and passes through a regime where it nearly vanishes. Following the steady-state PL analysis, we first integrate the two helicity-resolved spectra over the moiré emission band and then form the energy-integrated degree of circular polarization, $P_v$,

$$P_v = [\int_{\Delta E} (I\sigma^+(E) - I\sigma^-(E))\, dE] / [\int_{\Delta E} (I\sigma^+(E) + I\sigma^-(E))\, dE]$$

where $I\sigma^+$ and $I\sigma^-$ represent the PL intensities of the $\sigma^+$ and $\sigma^-$ helicity components in the common $\sigma^+$-excitation reference, and $\Delta E$ denotes the moiré emission band. This definition is equivalent to an intensity-weighted average of the energy-resolved DCP, not to an unweighted integral of $P(E)$. Because $P_v$ is an optical DCP, it is used here as a proxy for valley polarization rather than as a direct measure of a unique valley population imbalance. The gate-field dependence of $P_v$ along a diagonal dual-gate scan that varies carrier density at zero displacement field is summarized in Fig. 2(d) as a function of $V_b/t_b = V_t/t_t$, and the corresponding two-dimensional map plotted against $V_b/t_b$ and $V_t/t_t$ is shown in Fig. 2(e). Here, $t_t$ and $t_b$ denote the top and bottom hBN thicknesses, respectively. In principle, the observation of $P_v$ is not, by itself, sufficient to determine whether the valley polarization is rapidly lost or whether helicity-opposite components compensate after time integration. This distinction is particularly important in moiré heterobilayers, where different local registries can host excitonic states with different optical selection rules [10,15]. The steady-state data, therefore, motivate direct measurements of the time-domain helicity dynamics.

Figure 3(a) shows helicity-resolved time-resolved PL measured at a gate-field condition where the steady-state circular polarization is nearly zero. Despite the vanishing time-integrated polarization, the two helicity components are clearly different in the time domain. For the excitation-averaged traces, $\sigma^+$ and $\sigma^-$ are specified relative to a common $\sigma^+$-excitation reference (Sec. II of the Supplemental Material). In this convention, the $\sigma^-$ component is larger at early delay times, whereas the $\sigma^+$ component becomes larger at later delay times. As a result, the two traces cross at a finite delay time, denoted as $t_{cross}$. To emphasize this behavior, we plot the helicity-difference signal $D(t) = I\sigma^+(t) - I\sigma^-(t)$ in Fig. 3(b). With this definition, $D(t)$ is negative at early delay times and positive at later delay times. The signal, therefore, changes sign as a function of delay time, demonstrating that the apparent disappearance of the steady-state polarization results from temporal compensation between opposite helicity components. The corresponding time-dependent polarization, $P(t) = D(t)/S(t)$ ($S(t) = I\sigma^+(t) + I\sigma^-(t)$), is shown in Fig. 3(c). As clearly seen, a single-exponential depolarization curve cannot reproduce the observed sign reversal. This failure indicates that the data cannot be described by a single valley population that monotonically loses its polarization, and we need to explicitly account for the two contributions, which have different optical selection rules [10,15]. The sign reversal of $D(t)$ provides a direct way to understand the nearly zero steady-state polarization: the early- and late-time contributions have opposite signs, and integrating these negative and positive contributions over the emission decay drives the net helicity imbalance close to zero. Thus, the nearly vanishing steady-state $P_v$ does not imply that the valley polarization is absent at all delay times. Instead, it reflects a cancellation between temporally distinct helicity components. Additional gate-dependent decays, a comparison of steady-state and time-integrated $|P_v|$ maps, and excitation-helicity reversal are shown in Figs. S2–S4. The gate dependence clearly demonstrates that the spectral weight of competing two kinds of moiré excitons with opposite helicity can be controlled by gating.

We next introduce a minimal model to capture the helicity crossing. The model is motivated by A-like and B-like moiré emission channels with opposite circular optical selection rules [10,15]. Figure 4(a) schematically illustrates two local moiré registries that emit photons with opposite helicities, $\sigma^+$ and $\sigma^-$. We denote the corresponding helicity-opposite bright registries expected in R-type $WSe_2/WS_2$ as A and B [13,15]. The near-zero-DCP regions in Figs. 2(d,e) appear to be governed primarily by carrier density, consistent with this A/B interpretation. Because our experiment resolves emission dynamics rather than real-space site occupation, we refer to the corresponding components as A-like and B-like moiré emission channels. Spectrally isolating either of the two opposite-helicity channels would remove the interchannel compensation and helicity crossing described by this model. Such isolation is hindered by broad, overlapping emission features in our $WSe_2/WS_2$ device, in contrast to the well-resolved peaks reported in $MoSe_2/WSe_2$ [8,11]. The extracted $t_{cross}$ should not be interpreted as a microscopic A–B hopping time or a microscopic intervalley relaxation time; it is an emergent helicity-compensation time. Previously reported sub-picosecond site transitions in $WSe_2/WS_2$ precede interlayer-exciton formation and therefore do not directly determine the interchannel transfer rates relevant here [23]. Within the A/B interpretation, the reported resonance splitting (~50 meV [15]) suggests strongly suppressed thermally activated uphill transfer at 3.5 K ($k_BT \approx 0.30$ meV), although downhill or nonthermal transfer remains possible. As a minimal approximation neglecting interchannel transfer, we model the emission as two pre-established helicity channels that decay with distinct effective rates. The rate equations are

$$\mathrm{d}n_{i,+}/\mathrm{d}t = -\Gamma_i n_{i,+} - \gamma_i n_{i,+} + \gamma_i n_{i,-},$$

$$\mathrm{d}n_{i,-}/\mathrm{d}t = -\Gamma_i n_{i,-} - \gamma_i n_{i,-} + \gamma_i n_{i,+}.$$

Here, $n$ denotes the density of moiré excitons, and subscripts $i$ and ± represent the A/B moiré channels and K/−K valleys, respectively. $\Gamma_i$ and $\gamma_i$ denote the effective population decay rate and effective symmetric intervalley-scattering rate in each direction, respectively. The population and helicity-difference modes then decay at $\Gamma_i$ and $\lambda_i = \Gamma_i + 2\gamma_i$, respectively. The two-channel model is not intended to uniquely determine microscopic hopping pathways; rather, it provides the minimal phenomenology required by the observed sign reversal of $D(t)$. Because the A-like and B-like channels contribute with opposite circular selection rules, the experimentally measured helicity-difference signal is written as

$$D(t) = S(0)\mathrm{e}^{-\Gamma t}(D_A\, \mathrm{e}^{-2\gamma_A t} + D_B\, \mathrm{e}^{-2\gamma_B t}),$$

Here, $D_A$ and $D_B$ are dimensionless signed helicity weights normalized by $S(0)$, with $D_A D_B < 0$. For the analytic expression above, we assume a common population decay rate $\Gamma$ for the two channels; $\Gamma$ includes radiative decay, nonradiative decay, and possible unresolved relaxation processes within each emissive channel. This minimal two-channel expression naturally produces a zero crossing of $D(t)$ when the early-time and late-time helicity components have opposite signs. For $\gamma_A \neq \gamma_B$, the crossing time is given by

$$t_{cross} = [1/(2(\gamma_A - \gamma_B))] \ln(D_A/-D_B),$$

for $D_A D_B < 0$, provided $t_{cross} > 0$ and lies within the observation window. To limit the independently varied population parameters in the fit, we use a common envelope derived from the measured $S(t)$, writing $D(t) = S(t)[D_A \exp(-2\gamma_A t) + D_B \exp(-2\gamma_B t)]$, with $D_A D_B < 0$. The initial polarization magnitude is fixed to a common reference-derived value for both contributions and across the gate series; the fit conventions are given in the Supplemental Material. Figure 4(b) shows that this model reproduces the measured helicity-resolved dynamics, including the crossing between $\sigma^+$ and $\sigma^-$ emission. In this panel, the $\sigma^+$ component is plotted in blue and the $\sigma^-$ component in red. In contrast, a single-component depolarization model, $P(t) \propto \exp(-t/\tau_v)$, cannot generate a sign reversal of $D(t)$ without introducing an additional helicity-opposite component [6,20,21]. The minimal model therefore provides a compact phenomenological interpretation of the time-resolved data, rather than a unique microscopic fit: the nearly zero steady-state polarization arises from the competition between A-like and B-like moiré emission channels with opposite helicity projections and different effective relaxation dynamics [10,15]. This interpretation explains why a vanishing time-integrated valley polarization should not be equated with rapid valley depolarization. The present data do not reliably determine all component-specific population and valley lifetimes simultaneously. Distinct population lifetimes with a common intervalley-scattering rate can also reproduce the crossing and temporal compensation (Fig. S5). The A/B-site interpretation remains nonunique because singlet and triplet interlayer excitons with in-plane transition dipoles may also contribute [24] (see Sec. III of the Supplemental Material [25]).

Finally, we examine how the helicity-crossing dynamics evolve under electrostatic tuning. For the one-dimensional common-gate-field sweeps in Figs. 4(c) and 4(d), the common normalized gate coordinate is $V_b/t_b = V_t/t_t$; this configuration corresponds to applying the top and bottom gates with the same sign and is used to tune carrier injection. For each gate-field condition, we extract $t_{cross}$ from a local linear fit to $P(t)$ near its zero crossing when a clear sign reversal is observed; no $t_{cross}$ is assigned when $D(t)$ does not cross zero within the measured decay window. Figure 4(c) shows the resulting gate-field dependence of $t_{cross}$. The crossing time changes systematically with electrostatic tuning, indicating that the relative weights and effective relaxation rates of the helicity-opposite channels are electrically tunable [8,20]. A key observation is that finite $t_{cross}$ appears in the vicinity of the gate-field region where the time-integrated polarization vanishes. This correspondence shows that static polarization zero is not a featureless depolarized regime. Rather, it is a regime in which hidden time-domain helicity dynamics are especially important. Figure 4(d) shows the gate-field dependence of the fitted dimensionless weight magnitudes $|D_A|$ and $|D_B|$. These quantities are plotted as positive magnitudes for comparison, while $D_A$ and $D_B$ enter $D(t)$ with opposite signs in the model expression. The systematic gate-field dependence of $t_{cross}$ and of the extracted amplitude magnitudes distinguishes the observed behavior from accidental cancellation or measurement noise [8,19,20]. Within the minimal A/B-channel picture, changes in $t_{cross}$ can arise from changes in the relative initial amplitudes, effective lifetimes, or effective depolarization rates of the two helicity-opposite channels. Thus, the gate-field-dependent analysis provides information that is inaccessible from steady-state polarization alone. It reveals that the valley dynamics hidden behind a vanishing integrated polarization remain well defined and can be electrically controlled.

In summary, we have shown that a vanishing steady-state valley polarization in electrically tunable $WSe_2/WS_2$ moiré excitons can conceal well-defined time-domain helicity dynamics. Helicity-resolved time-resolved PL reveals a temporal crossing between $\sigma^+$ and $\sigma^-$ emission components in the regime where the time-integrated polarization nearly vanishes, demonstrating that the apparent polarization cancellation does not correspond to a featureless depolarized state. A minimal A/B-like two-channel model, in which helicity-opposite moiré emission channels share a common population envelope derived from the measured total intensity but possess different effective valley depolarization rates and gate-dependent amplitudes, captures the observed crossing. The systematic gate-field dependence of $t_{cross}$ is therefore interpreted as electrostatic tuning of the relative helicity-channel weights and effective dynamics, rather than as a direct measure of microscopic A–B hopping or a single intrinsic valley lifetime [6,20,21,23]. More generally, these results show that time-integrated circular polarization can give a false negative for valley polarization whenever multiple helicity channels with different decay kinetics coexist. They establish time-resolved helicity measurements as a sensitive probe of hidden-valley dynamics that are inaccessible via steady-state polarization alone.

## SUPPLEMENTAL MATERIAL

See the Supplemental Material [25] for the SHG characterization used to determine the twist angle and stacking configuration and for the analysis conventions used to define $D(t)$, evaluate the time-integrated TRPL polarization, extract $t_{cross}$, and report the two-channel fit amplitudes. Additional optical conditions, polarization controls, model comparisons, and crossing-time uncertainty are also provided.

## ACKNOWLEDGMENTS

R.K. was supported by JSPS KAKENHI Grant No. JP25K22210, JP24H02218, and JP23H05469, and the World Premier International Research Center Initiative (WPI), MEXT, Japan. D.K. was supported by JSPS KAKENHI Grant No. JP26K22704; Canon Foundation; Murata Science and Education Foundation; Shorai Foundation for Science and Technology.

## CONFLICT OF INTEREST

The authors have no conflicts to disclose.

## AUTHOR CONTRIBUTIONS

Yuto Urano: Data curation (lead); Methodology (equal). Lata Chouhan: Data curation (supporting). Nurul Fariha Ahmad: Data curation (supporting). Kenji Watanabe: Resources (equal); Writing – review & editing (supporting). Takashi Taniguchi: Resources (equal); Writing – review & editing (supporting). Daichi Kozawa: Writing – review & editing (supporting). Ryo Kitaura: Conceptualization (lead); Supervision (lead); Data curation (equal); Methodology (equal); Writing – original draft (lead); Writing – review & editing (lead).

## DATA AVAILABILITY

The data that support the findings of this study are available from the corresponding author upon reasonable request.

## FIGURES AND CAPTIONS

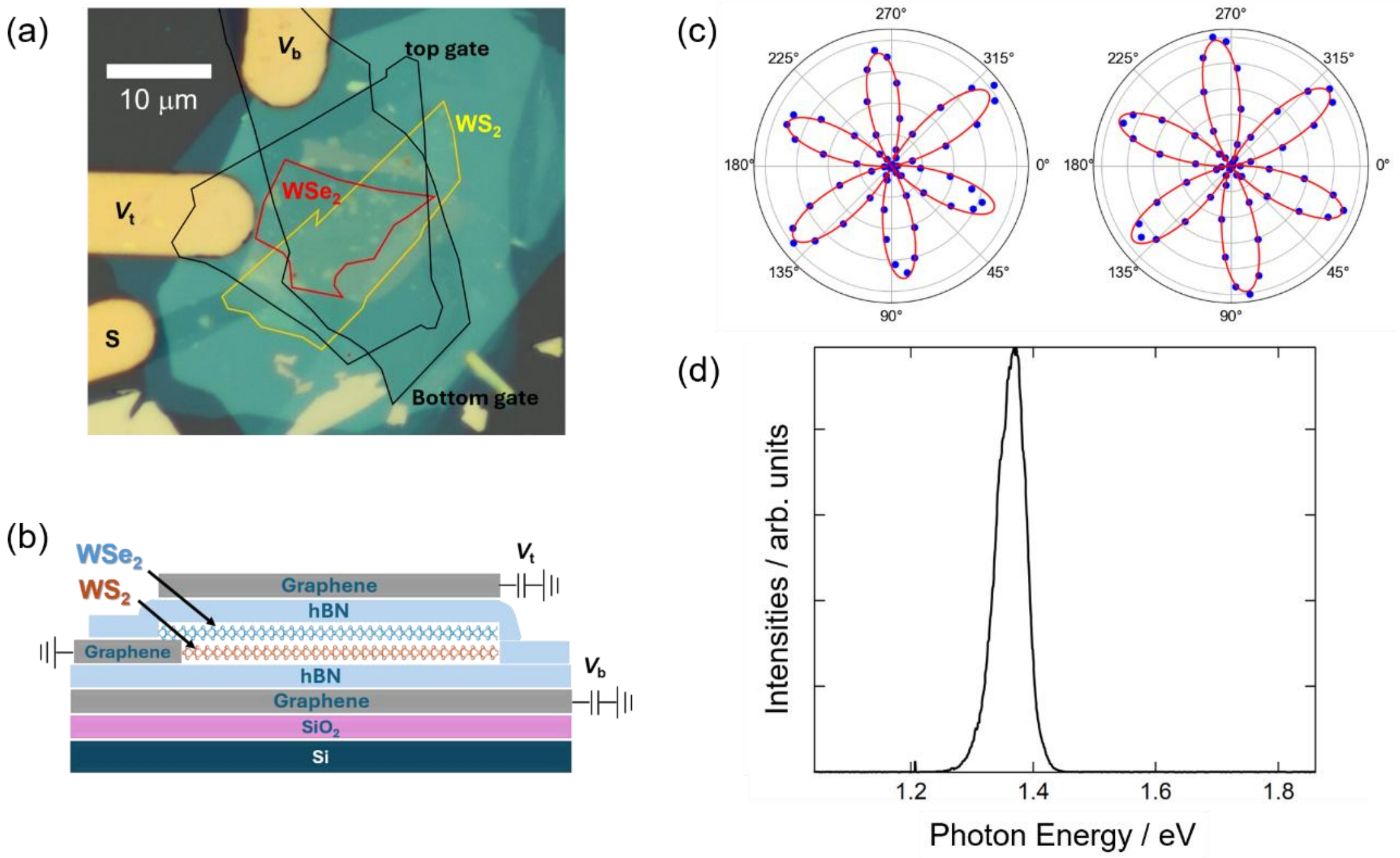

**FIG. 1. Device structure and optical characterization of the $WSe_2/WS_2$ heterostructure.** (a) Optical micrograph of the dual-gated $WSe_2/WS_2$ heterostructure device. The region used for helicity-resolved PL measurements is indicated. (b) Schematic cross-section of the device. (c) SHG characterization of the constituent monolayers, $WSe_2$ (left) and $WS_2$ (right), yielding a relative twist angle $\theta = 0.14 \pm 0.20°$. The heterobilayer SHG response is shown in Fig. S1 of the Supplemental Material; its enhanced intensity indicates constructive interference and verifies the near-0° (R-type) stacking configuration of the heterobilayer. (d) Typical PL spectrum measured at 4 K with $V_t = V_b = 0$, showing the dominant interlayer-exciton emission and the absence of detectable intralayer-exciton emission.

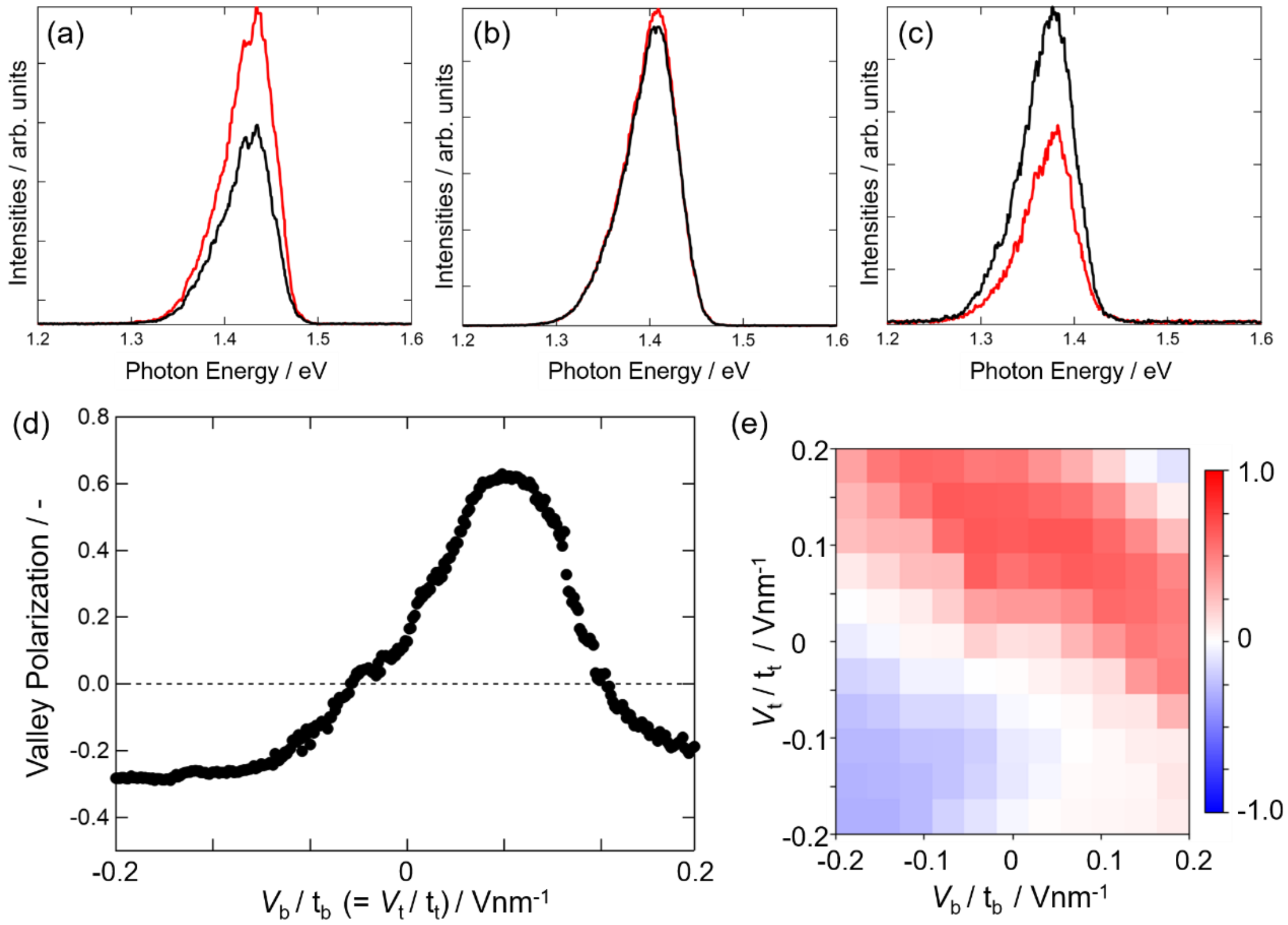

**FIG. 2. Gate-field-controlled steady-state valley polarization.** (a–c) Helicity-resolved photoluminescence spectra measured at representative gate-field conditions: (a) $V_b/t_b = 0.20$ V nm$^{-1}$ and $V_t/t_t = -0.08$ V nm$^{-1}$, (b) $V_b/t_b = 0.08$ V nm$^{-1}$ and $V_t/t_t = -0.08$ V nm$^{-1}$, and (c) $V_b/t_b = -0.20$ V nm$^{-1}$ and $V_t/t_t = -0.08$ V nm$^{-1}$. These conditions correspond to positive, nearly zero, and negative time-integrated circular polarization, respectively. Using the $\sigma^+$-excitation reference, red and black curves correspond to $\sigma^+$ and $\sigma^-$ emission, respectively. (d) Gate-field dependence of integrated valley polarization plotted as a function of $V_b/t_b = V_t/t_t$ along a diagonal dual-gate scan at zero displacement field, showing a continuous evolution through zero. (e) Two-dimensional map of integrated valley polarization plotted against $V_b/t_b$ and $V_t/t_t$. The contour where valley polarization becomes zero marks the regime in which steady-state measurements alone cannot determine the underlying valley dynamics.

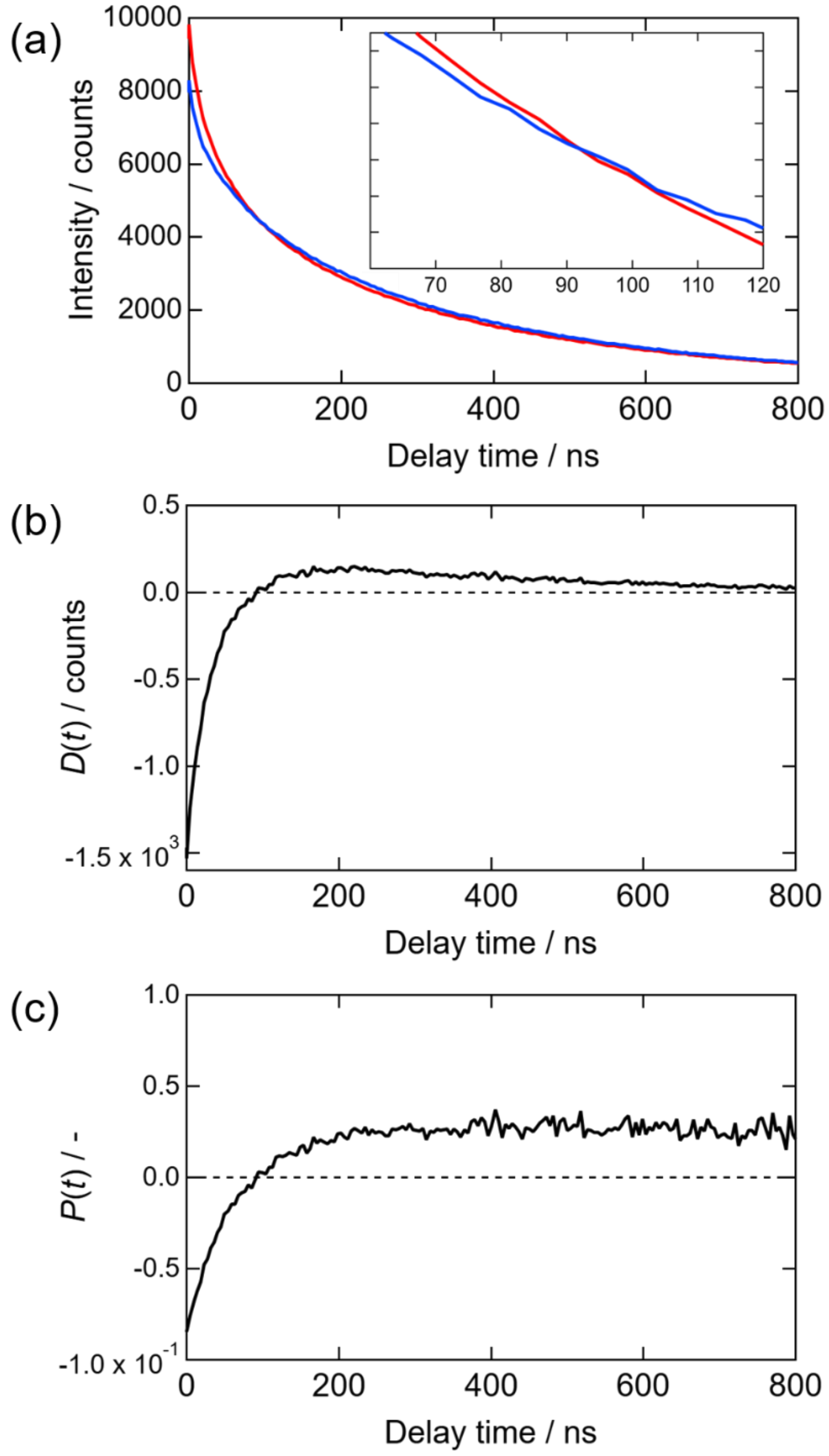

**FIG. 3. Time-resolved helicity dynamics at vanishing steady-state polarization.** (a) Time-resolved $\sigma^+$ (blue) and $\sigma^-$ (Red) PL measured at $V_b/t_b = V_t/t_t$ = -0.037 V nm$^{-1}$, where $P_v$ is close to 0. Under the common excitation reference defined in Sec. II of the Supplemental Material, $\sigma^-$ is larger at early delays and $\sigma^+$ becomes larger at later delays, producing a finite crossing time $t_{cross}$. The inset shows an enlarged view of the 60-120 ns cutout. (b) Helicity-difference signal $D(t) = I\sigma^+(t) - I\sigma^-(t)$, showing a negative-to-positive sign reversal in the time domain. (c) Time-dependent polarization $P(t) = D(t)/S(t)$. A single-exponential depolarization curve fails to reproduce the observed dynamics.

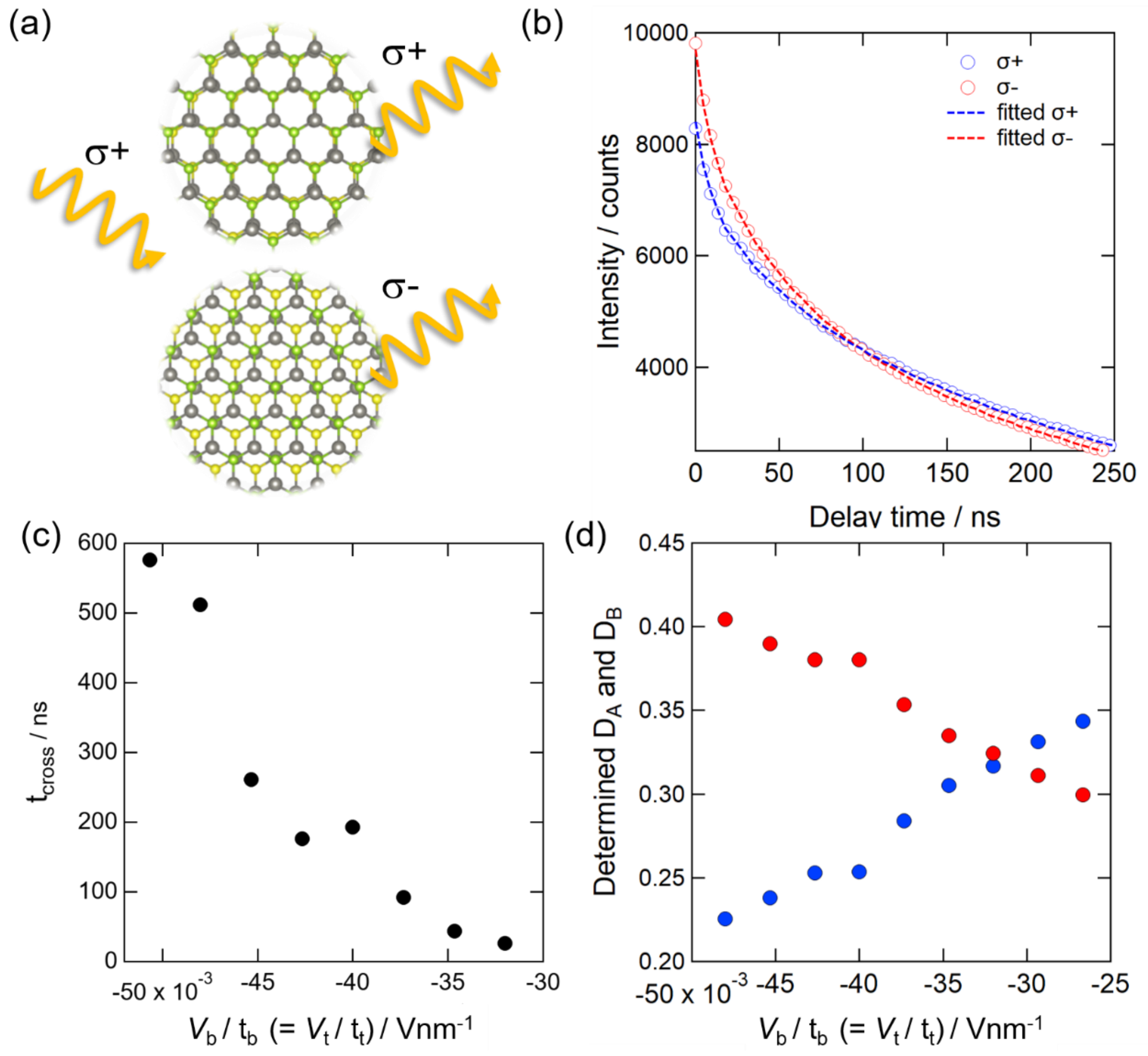

**FIG. 4. Minimal A/B-channel model for helicity crossing.** (a) Schematic illustration of one possible microscopic assignment to two local moiré registries with opposite circular optical selection rules, corresponding to $\sigma^+$ and $\sigma^-$ emission channels [10,15]. (b) Fit of the time-resolved helicity dynamics using the minimal model with a common measured population envelope. Blue and red symbols denote $\sigma^+$ and $\sigma^-$ emission, respectively, and dashed curves show the model fits. The crossing is reproduced without assuming a single rapid valley depolarization process. The time origin in (b,c) is set at the PL rise. (c) Gate dependence of $t_{cross}$, estimated from local linear fits to $P(t)$ near zero along $V_b/t_b = V_t/t_t$ (Sec. VI of the Supplemental Material). (d) Common-gate-field dependence of the fitted dimensionless weight magnitudes $|D_A|$ and $|D_B|$. The plotted quantities are positive magnitudes (blue: $|D_A|$; red: $|D_B|$); in the model expression, $D_A$ and $D_B$ have opposite signs.

Supplemental Material for

# Hidden valley dynamics behind vanishing circular polarization in moiré excitons

*Yuto Urano[1,2], Lata Chouhan[1], Nurul Fariha Ahmad[1], Kenji Watanabe[3], Takashi Taniguchi[1], Daichi Kozawa[1,4], and Ryo Kitaura[1,2,*]*

[1] *Research Center for Materials Nanoarchitectonics, National Institute for Materials Science, 1-1 Namiki, Tsukuba, Ibaraki 305-0044, Japan*

[2] *Graduate School of Chemical Sciences and Engineering, Hokkaido University, 5, Kita 8 Nishi, Kita-ku, Sapporo-city, Hokkaido 060-8628, Japan*

[3] *Research Center for Electronic and Optical Materials, National Institute for Materials Science, 1-1 Namiki, Tsukuba, Ibaraki 305-0044, Japan*

[4] *Department of Materials Science, Institute of Pure and Applied Sciences, University of Tsukuba, 1-1-1 Tennodai, Tsukuba, Ibaraki 305-8573, Japan*

* Correspondence to KITAURA.Ryo@nims.go.jp

## I. SHG CHARACTERIZATION

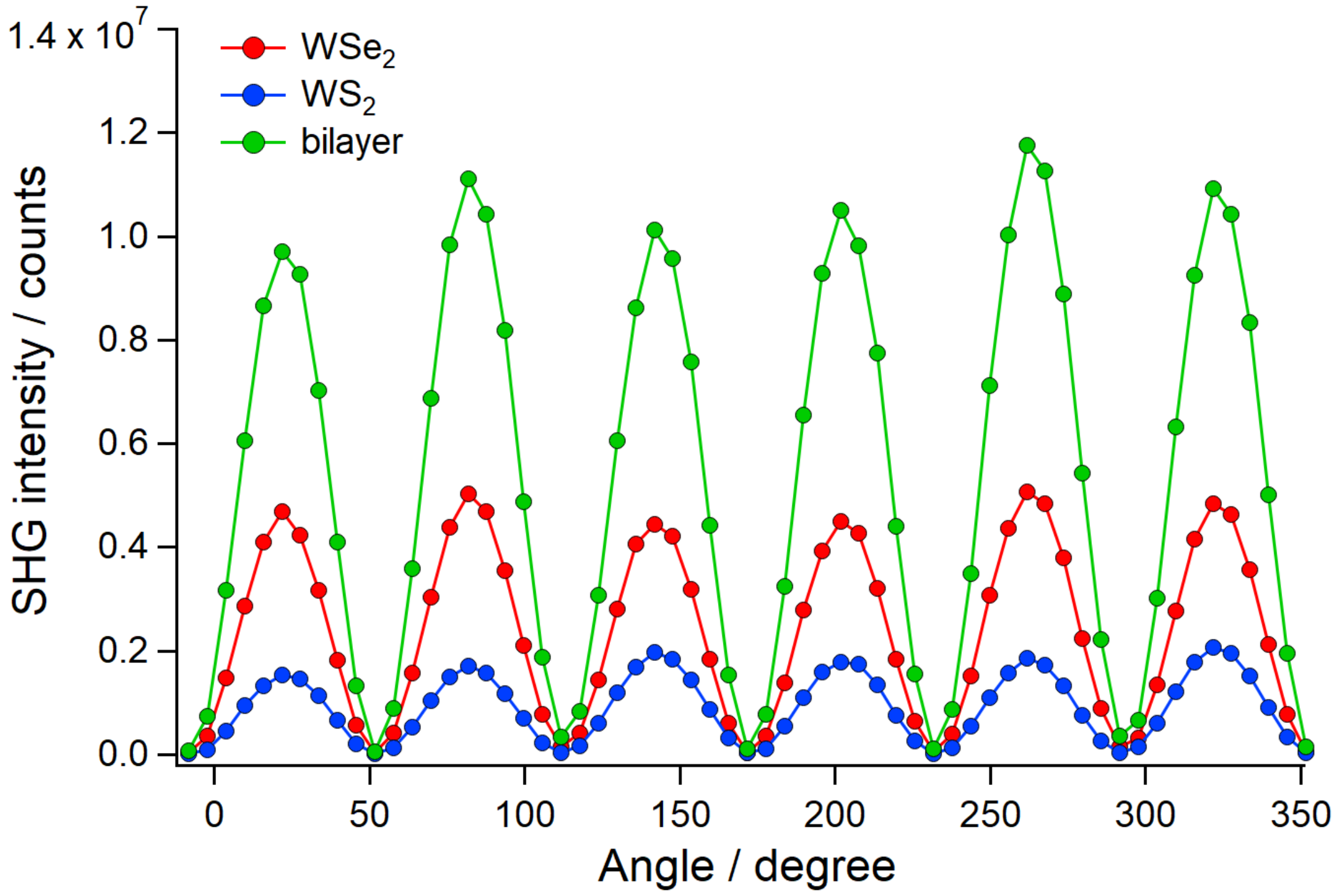

**FIG. S1.** Angle-dependent SHG intensity of the $WSe_2/WS_2$ heterostructure. SHG intensity measured from the $WS_2$ monolayer, $WSe_2$ monolayer, and $WSe_2/WS_2$ heterobilayer region as a function of polarization angle. The armchair orientations extracted from the monolayer regions are $-0.43 \pm 0.17°$ for $WS_2$ and $-0.57 \pm 0.10°$ for $WSe_2$, corresponding to a relative twist angle $\theta = 0.14 \pm 0.20°$. The enhanced SHG response from the heterobilayer region indicates constructive interference of the SHG fields, confirming the near-0° (R-type) stacking configuration.

## II. ANALYSIS CONVENTIONS

Opposite excitation helicities were prepared by rotating a half-wave plate in the excitation path, and both p and s detection channels were recorded for each setting. After background subtraction, we formed $p_+ + p_-$ and $s_+ + s_-$, where subscripts denote excitation helicity. Relative detector gain and timing offset were fitted jointly to match these summed transients, using the same correction for both excitation settings. Below, $p_\pm(t)$ and $s_\pm(t)$ denote corrected signals, with s and p corresponding to detected $\sigma^+$ and $\sigma^-$, respectively. With $D_\pm(t) = s_\pm(t) - p_\pm(t)$, we extract the excitation-odd component $D(t) = [D_+(t) - D_-(t)]/2$. Sum matching suppresses the excitation-even difference $[D_+(t) + D_-(t)]/2$, and the corrected traces show opposite helicity ordering (Fig. S4). We use a common $\sigma^+$-excitation reference, with $I_{\sigma+}$ and $I_{\sigma-}$ denoting the corrected co- and cross-circular intensities. The fixed-port traces in Fig. S2 use the same reference. The helicity-difference signal is defined as $D(t) = I\sigma^+(t) - I\sigma^-(t)$. With this convention, the representative data in Fig. 3 have a negative early-time contribution and a positive late-time contribution; the small time-integrated polarization reflects compensation between these temporally distinct contributions. In Fig. S4, the two excitation settings are shown separately, with the excitation and detected helicities specified individually.

The time-integrated TRPL polarization is evaluated by first integrating the two helicity-resolved intensities over the measured decay and then forming $P_{\mathrm{TRPL}} = [\int(I\sigma^+(t) - I\sigma^-(t))\mathrm{d}t]/[\int(I\sigma^+(t) + I\sigma^-(t))\mathrm{d}t]$. This definition matches the steady-state PL analysis and is not an unweighted time integral of $P(t)$. The crossing time $t_{\mathrm{cross}}$ is assigned only when $D(t)$ changes sign within the measured delay window and is obtained from a local linear fit to $P(t) = D(t)/S(t)$ near zero (Sec. VI). Traces without a sign reversal are not assigned a crossing time. $D_\mathrm{A}$ and $D_\mathrm{B}$ are dimensionless signed weights in $D(t) = S(t)[D_\mathrm{A} \exp(-2\gamma_\mathrm{A} t) + D_\mathrm{B} \exp(-2\gamma_\mathrm{B} t)]$, with $D_\mathrm{A} D_\mathrm{B} < 0$. Figure 4(d) shows $|D_\mathrm{A}|$ in blue and $|D_\mathrm{B}|$ in red. In the fit convention of Sec. V, $D_\mathrm{A} = P_0 f_\mathrm{A} \exp(-\kappa_\mathrm{A} t_\mathrm{d})$ and $D_\mathrm{B} = -P_0(1 - f_\mathrm{A})\exp(-\kappa_\mathrm{B} t_\mathrm{d})$. Thus, $D(0) = S(0)(D_\mathrm{A} + D_\mathrm{B})$; the fixed $P_0$ applies at $t = -t_\mathrm{d}$ rather than fixing $|D_\mathrm{A}| + |D_\mathrm{B}|$ at $t = 0$.

## III. MEASUREMENT CONDITIONS

PL was collected using an NA = 0.42 objective (LCD Plan Apo NIR 50, Mitutoyo), and the entire collected PL passed through a long-pass filter (FELH0750, Thorlabs). After separation of the p- and s-polarized components by a polarizing beam splitter, each component was sent to a separate single-photon avalanche photodiode (SPCM-900-14-FC, Excelitas) for simultaneous detection. A quarter-wave plate (AQWP05M-980, Thorlabs) placed immediately before the objective produced circular excitation and converted the returning circular components into orthogonal linear polarizations. The acquisition of the two excitation-helicity datasets, relative detector correction, and extraction of $D(t)$ are described in Sec. II. Using this setup, for ideal collection centered on the sample normal, a purely out-of-plane dipole produces radial polarization with equal circular projections. This contribution adds equally to the two detected intensities and can dilute $P(t)$, but cannot by itself reverse $D(t)$. This argument does not exclude spin-triplet states with in-plane dipoles or other spin-dependent emission channels.

Time-resolved measurements were performed with a 705-nm pulsed diode laser (LDH-*D*-C-705, PicoQuant) delivering approximately 70 ps pulses at 500 nW, with a 2-μs repetition period (0.5 MHz). Assuming a uniformly filled circular pupil, $\lambda$ = 705 nm and NA = 0.42 yield a nominal Airy-disk diameter (between first minima) of $1.22\lambda/\text{NA} \approx 2.0$ μm and a Rayleigh lateral resolution of $0.61\lambda/\text{NA} \approx 1.0$ μm. The measured instrument-response function (IRF) had a full width at half maximum (FWHM) of 557 ps. Photon arrival times were recorded using a Time Tagger Ultra (Swabian Instruments) with a 200-ps acquisition-bin width. The excitation was approximated as a delta function; the crossing-time analysis is described in Sec. VI. For an exponential PL population-decay component, the preceding-pulse contribution relative to the current pulse is $\exp(-T/\tau)/[1 - \exp(-T/\tau)]$. For $T = 2$ μs and the representative population-decay component with $\tau \approx 0.37$ μs, this ratio is approximately 0.5%, indicating a small correction for that component.

## IV. ADDITIONAL POLARIZATION MEASUREMENTS

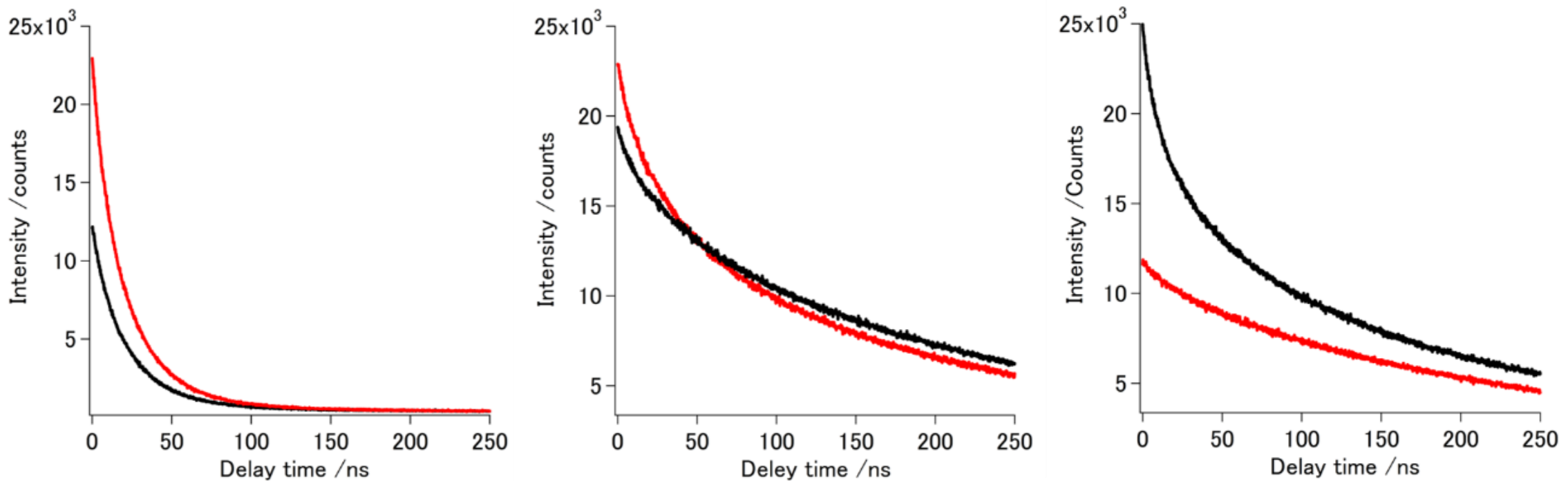

FIG. S2. PL decays resolved into $\sigma^+$ (red) and $\sigma^-$ (black) components, shown using the common $\sigma^+$-excitation reference defined in Sec. II, at the conditions of main-text Figs. 2(a), 2(b), and 2(c), from left to right.

The helicity ordering reverses between the gate conditions. At the intermediate condition, the traces cross, and their early- and late-time intensity differences compensate, hiding the dynamics in the nearly zero integrated DCP.

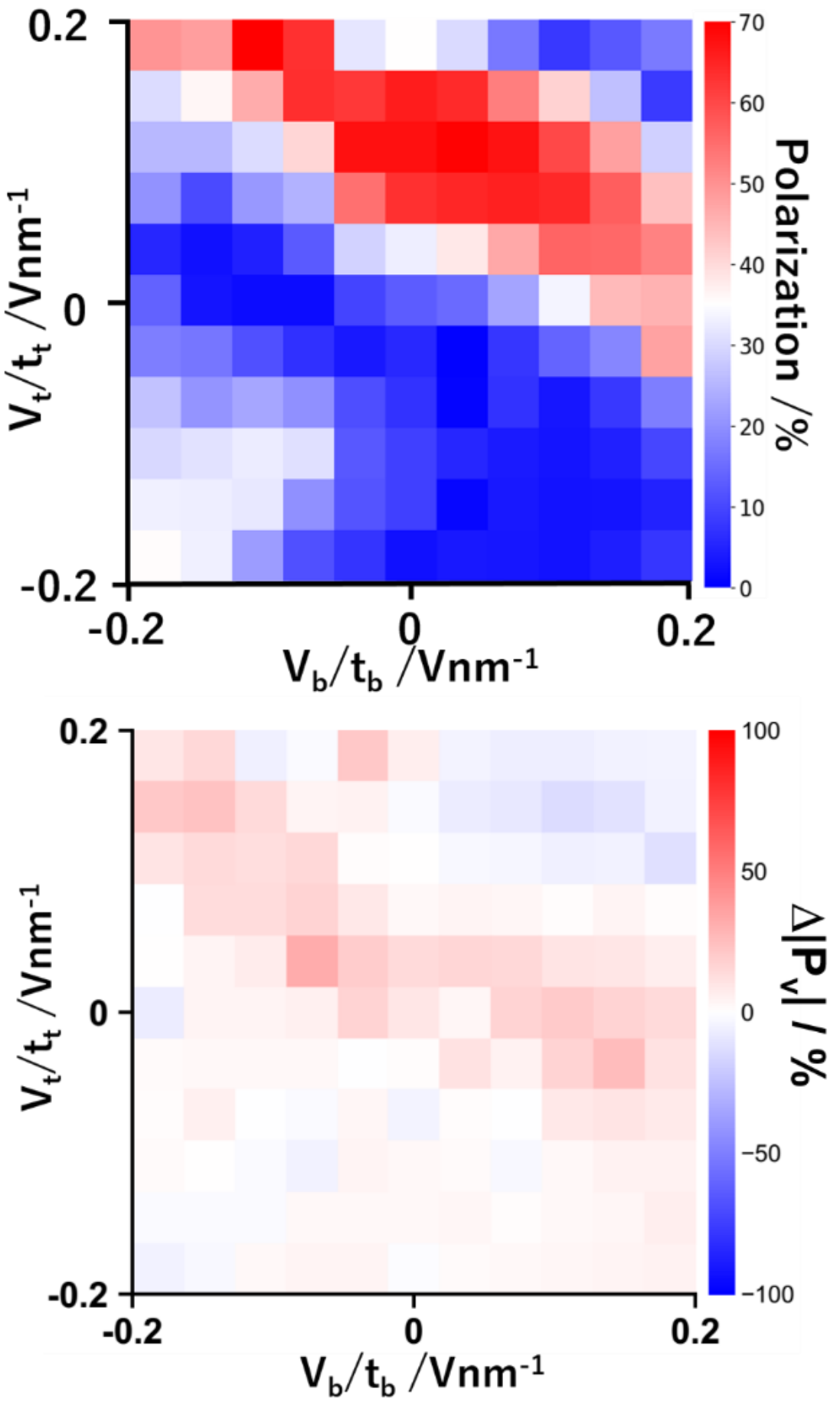

FIG. S3. Circular-polarization magnitude $|P_v|$ from steady-state polarization-resolved PL (PRPL, upper left) and time-integrated time-resolved polarization-resolved PL (TRPRPL, upper right). The TRPRPL intensities are integrated before forming their normalized difference, as defined in Sec. II. Both upper maps show $|P_v|$ in percent. The lower-left map shows the TRPRPL magnitude minus the PRPL magnitude in percentage points, with a color scale from −100 to +100. The integration window and background treatment are the same as for the other time-resolved measurements. Measurement conditions are given in Sec. III.

Both measurements reproduce the same overall gate-dependent regions of suppressed and enhanced polarization, supporting consistency between steady-state PL and time-integrated TRPL. Across the 121 corresponding gate points, the mean absolute difference is 7.01 percentage points and the root-mean-square difference is 9.41 percentage points. These differences quantify intermeasurement deviations; pointwise statistical uncertainties cannot be inferred from the polarization maps alone.

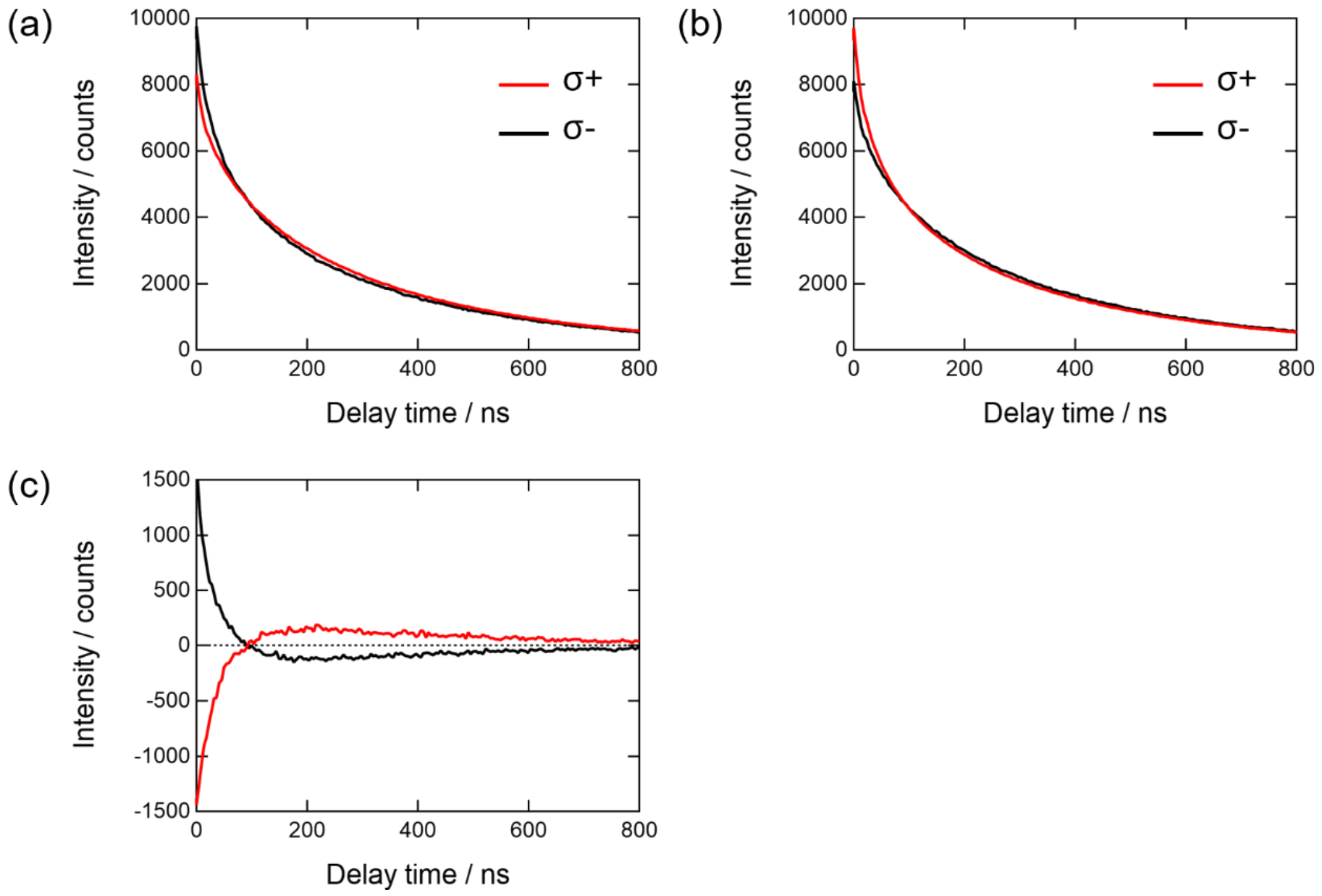

FIG. S4. Excitation-helicity reversal of time-resolved PL. (a,b) Detected $\sigma^+$ (red) and $\sigma^-$ (black) PL under $\sigma^+$ and $\sigma^-$ excitation, respectively. (c) Helicity differences $D_{\pm}(t) = I_{\sigma+}(t) - I_{\sigma-}(t)$ under $\sigma^+$ (red) and $\sigma^-$ (black) excitation. The intensities are shown after correction for relative detection efficiency and timing. Measurement conditions are given in Sec. III; the detector correction and definitions of $D_{\pm}(t)$ and $D(t)$ are described in Sec. II.

Reversing the excitation helicity reverses the early- and late-time helicity differences, while a temporal crossing remains visible for both settings. This supports the helicity-dependent character of the dynamics. The approximate reversal symmetry after sum matching is not an independent validation of the detector calibration (Sec. II).

## V. MODEL COMPARISON

We use constrained effective models because simultaneous estimation of all component-specific population and valley lifetimes is unstable. The main fit in Fig. 4(b) uses a common population envelope $S_{\mathrm{env}}(t)$ derived from the measured total PL intensity. The weights in Fig. 4(d) are obtained by fitting $D(t)$. We obtain this envelope by a least-squares adjustment that enforces a nonincreasing time dependence, with any negative values set to zero. Its polarization is $P(t) = P_0[f_A \exp(-\kappa_A(t + t_d)) - (1 - f_A)\exp(-\kappa_B(t + t_d))]$, with $D(t) = S_{\mathrm{env}}(t)P(t)$. Here $t$ is the delay time measured from the first fitted point ($t = 0$), $t_d = 5$ ns, and $P_0 \approx 0.654$ is fixed from reference measurements for both contributions across the gate series. Thus, $P_0$ refers to the extrapolated coordinate $t = -t_d$. The fitted $\kappa_i$ describe conditional polarization lifetimes $\tau_{\mathrm{pol},i} = 1/\kappa_i$; for symmetric intervalley scattering, $\kappa_i = 2\gamma_i$. The fit varies $f_A$ and two positive polarization lifetimes, with $0 \leq f_A \leq 1$. Because correlations between fit parameters can lead to unstable estimates, we apply a soft constraint. The penalty is applied only when the larger polarization lifetime exceeds five times the smaller, using the additional residual $\sqrt{50}\max[0, |\ln \tau_{\mathrm{pol,A}} - \ln \tau_{\mathrm{pol,B}}| - \ln 5]$. Larger ratios remain allowed; the fitted ratio in Fig. 4(b) is approximately 1.56, so this residual is zero. As alternatives, independent channels give $S(t) = \Sigma_{\mathrm{i}}\, a_i \exp(-\Gamma_i\, t)$ and $D(t) = \Sigma_{\mathrm{i}}\, d_i \exp[-(\Gamma_i + 2\gamma_i)t]$, with $a_i \geq |d_i|$, opposite signs of $d_i$, and nonnegative rates. The two population amplitudes, population lifetimes, and initial component polarizations are free. A common or separate $\gamma$ gives seven or eight parameters. For the fit in Fig. 4(b), the difference between calculated and measured intensity at each time point is divided by the maximum measured intensity across both helicity channels. These normalized residuals and the lifetime-ratio penalty are evaluated using a soft-L1 loss, which reduces the influence of large residuals compared with ordinary least squares. For the alternative models, we minimize the sum of squared intensity residuals, each multiplied by $5/\max(I, 1)$, where $I$ is the measured intensity. This weighting gives weaker signals greater relative weight. All three stored fits are retained unchanged and use the same 276 points at 4.5-ns spacing over 0–1237.5 ns; restricting the display to 0–400 ns does not change the fitting window. This comparison is descriptive because the constraints and fitting objectives differ.

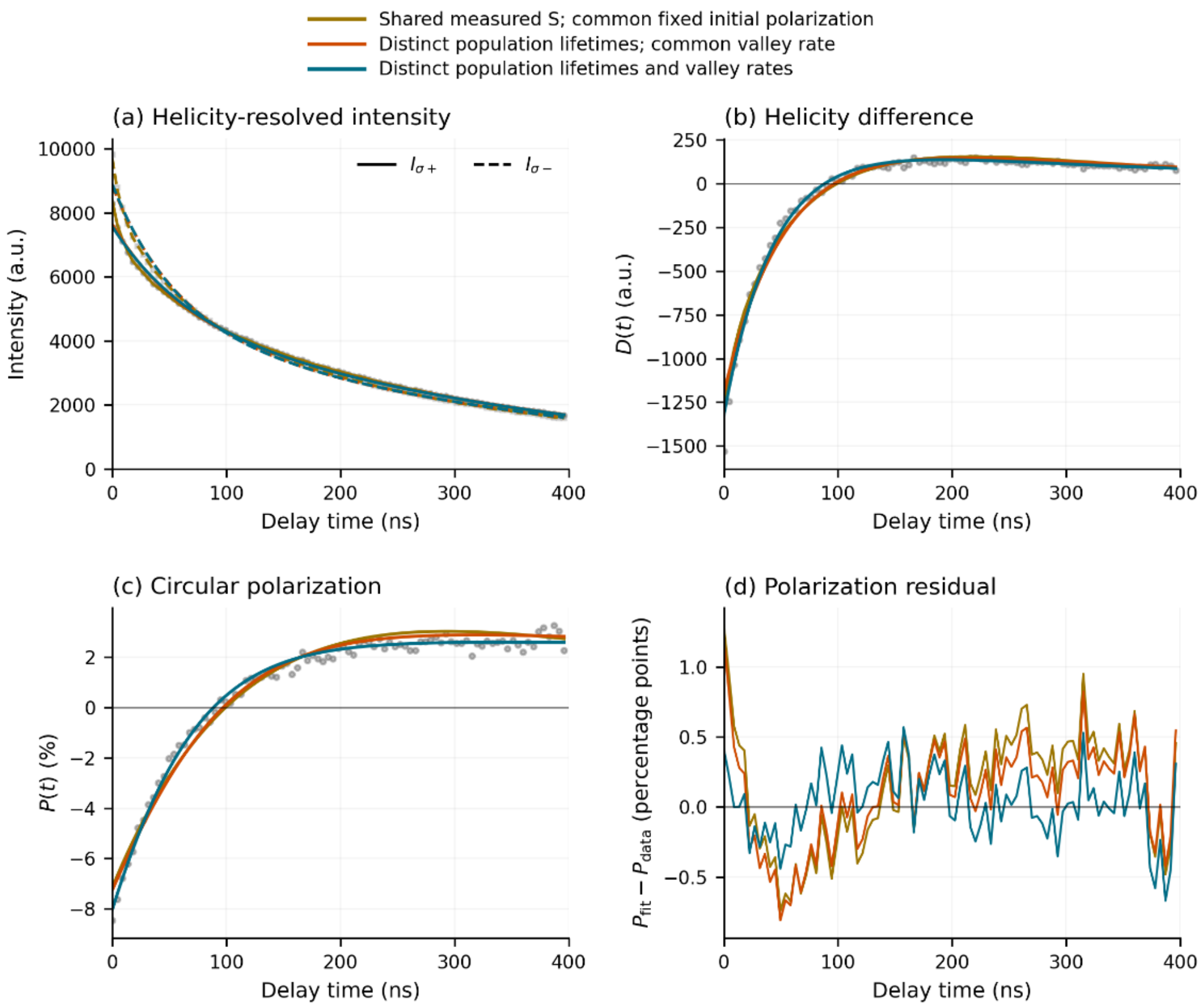

FIG. S5. Representative fit comparison: (a) detected intensities on a linear scale, (b) $D(t)$, (c) $P(t)$, and (d) polarization residuals. Gray markers are measurements. Gold: measured-envelope fit with fixed $P_0$; orange: distinct population lifetimes with common $\gamma$; blue: distinct population lifetimes and $\gamma$. All panels display 0–400 ns from the first fitted point.

All three models reproduce the principal helicity evolution and crossing over the displayed 0–400 ns, although their detailed residuals differ. Distinct population lifetimes with a common intervalley-scattering rate provide an alternative explanation of the crossing and compensation. However, the inferred valley-relaxation times depend strongly on the model assumptions; the present data do not reliably determine all component-specific population and valley lifetimes simultaneously. We therefore retain the constrained measured-envelope description in the main text. A closer match to a small polarization at long delays does not by itself establish a preferred microscopic mechanism. The consistent conclusion is that opposite-helicity contributions with different temporal evolutions can account for the observed crossing and compensation.

TABLE S1. Conditional fit parameters for Fig. S5. Negative and positive refer to signs of contributions to $D$, not established A/B sites. For the measured-envelope fit, $\gamma$ denotes $\kappa/2$, corresponding to polarization time constants of 157 and 244 ns; no exponential population lifetime is assigned to this envelope.

| Constraint | Population lifetime negative / positive (ns) | $\gamma$ negative / positive ($\mu s^{-1}$) |
|---|---|---|
| Common measured envelope | Measured $S(t)$ | 3.19 / 2.05 |
| Different populations, common $\gamma$ | 56.1 / 368 | 0.321 / 0.321 |
| Separate populations and $\gamma$ | 56.6 / 368 | 2.64 / 0.085 |

For the retained measured-envelope fit in Table S1, a local linearization of the 276 $D(t)$ samples at the stored solution gives conditional 1σ sensitivity scales of 0.066 and 0.050 $\mu s^{-1}$ for the negative- and positive-contribution rates, respectively, with a correlation of +0.963. This calculation fixes $P_0$, $S_{env}(t)$ and $t_d$ and assumes independent, equal-variance residuals. These local diagnostics exclude temporal correlations and uncertainty in the fixed quantities and detector corrections; they are not the full standard errors of the original constrained fit. The common-rate alternative gives $\gamma = 0.321\ \mu s^{-1}$ with a conditional local standard error of 0.087 $\mu s^{-1}$. The fully independent rates are $2.64 \pm 0.79$ and $0.085 \pm 0.106\ \mu s^{-1}$ for negative and positive contributions, respectively. The positive rate is therefore weakly constrained. For positive contributions, amplitude–lifetime correlations are −0.91 for the population component and −0.89 for the helicity-difference component. These local estimates assume the selected model and relative independent-error weights; they are not calibrated confidence intervals.

## VI. CROSSING-TIME ESTIMATION

The crossing times in Fig. 4(c) of the main text are estimated from the measured polarization $P(t) = D(t)/S(t)$, formed from the same detector-corrected, excitation-averaged and background-processed intensities used in Fig. 4(b), independently of the decay-model fits. The preprocessing subtracts $-[D_{+}(t) + D_{-}(t)]/2$ from both intensities, removes each channel's pre-rise baseline over 30–50 ns, and symmetrically removes the mean late-time difference over 1700–1800 ns; these intervals refer to recorded time. Because $S(t)$ is positive in the selected windows, $P(t)$ and $D(t)$ have the same zero crossing. Here $t$ is measured from the PL rise (approximately 60 ns on the recorded time axis). A line $P(t) = mt + b$ is fitted to ten consecutive samples after the PL rise, at 4.5-ns spacing (a 40.5-ns window span). The first window whose measured values bracket zero and whose fitted root lies within the window is retained, giving $t_{\text{cross}} = -b/m$.

For a fixed-window ordinary least-squares fit with independent equal-variance residuals, $C = [\text{RSS}/(N-2)](X^{\text{T}} X)^{-1}$ is the slope–intercept covariance, where RSS is the residual sum of squares of the local linear fit, $N$ is the number of samples in the fit window ($N = 10$ for the reported crossing times), and $X$ has columns $t$ and 1. Propagation gives $\text{Var}(t_{\text{cross}}) = g^{\text{T}} Cg$, with $g = (b/m^2, -1/m)$. For the eight points in Fig. 4(c) of the main text, ordered from left to right, the propagated fitting errors are 11.51, 3.06, 7.08, 1.97, 2.94, 1.73, 0.73, and 0.94 ns. For the representative sixth point, changing the window size to eight or twelve samples changes the estimate over a 1.2-ns range. These local, linearized 1σ errors are conditional on the selected windows and assumed independent, equal-variance residuals; they exclude systematic uncertainty.